\documentstyle[12pt]{article}
\newcommand{\mH}{M_{H}}

\newcommand{\mZ}{M_{Z}}

\newcommand{\signal}{p \bar{p} \to W H X}
\newcommand{\proda}{p \bar{p} \to H X}
\newcommand{\prodb}{p \bar{p} \to  t \bar{t} H X}

\newcommand{\bgA}{p \bar{p}  \to W b \bar{b} X }
\newcommand{\bgB}{p \bar{p} \to W Z X }
\newcommand{\bgC}{p \bar{p} \to W j j X}
\newcommand{\bgD}{p \bar{p} \to t \bar{t} X }

\newcommand{\be}{\begin{equation}}
\newcommand{\ee}{\end{equation}}
\newcommand{\bea}{\begin{eqnarray}}
\newcommand{\eea}{\end{eqnarray}}

\def\signature{``$e^\pm (\mu^\pm) + 2$ bottom jets''{}}

\def\eg{\hbox{\it e.g.}{}}

\relax

\def\figcap{\section*{Figure Captions\markboth
     {FIGURECAPTIONS}{FIGURECAPTIONS}}\list
     {Fig. \arabic{enumi}:\hfill}{\settowidth\labelwidth{Fig. 999:}
     \leftmargin\labelwidth
     \advance\leftmargin\labelsep\usecounter{enumi}}}
 \relax
\def\reflist{\section*{REFERENCES\markboth
     {REFLIST}{REFLIST}}\list
     {[\arabic{enumi}]\hfill}{\settowidth\labelwidth{[999]}
     \leftmargin\labelwidth
     \advance\leftmargin\labelsep\usecounter{enumi}}}
 \relax
\def\tabcap{\section*{Tables\markboth
     {TABLES}{TABLES}}\list
     {Table \arabic{enumi}:\hfill}{\settowidth\labelwidth{Table 999:}
     \leftmargin\labelwidth
     \advance\leftmargin\labelsep\usecounter{enumi}}}
 \relax

\begin{document}
\begin{titlepage}
 \null
 \vskip 0.5in
\begin{center}
 \makebox[\textwidth][r]{IP/BBSR/2000-10}

 \vspace{.15in}
  {\Large
     Prospects of Discovering the Higgs Boson at the Tevatron 
    }
  \par
 \vskip 1.5em
 {\large
  \begin{tabular}[t]{c}
   Pankaj Agrawal \\
\em Institute of Physics \\
\em Sachivalaya Marg \\
 \em Bhubaneswar, Orissa 751005 India\\
  \end{tabular}}
 \par \vskip 5.0em
 {\large\bf Abstract}
\end{center}
\quotation

  We carry out an event generator based study
 to evaluate the usefulness
 of the the exclusive signature ``$e^\pm (\mu^\pm) + 2$ bottom
 jets'' in discovering the Higgs boson at the Tevatron. 
 Using an event generator, we show that with enough
 luminosity, one should be able to detect the Higgs boson
 up to the mass of about 130 GeV. We consider a slightly modified 
 version of the above signature that includes one extra jet.
 This new signature is more useful because it increases the
 number of signal events significantly.

\endquotation
\vspace{1.0in}
 \baselineskip 10truept plus 0.2truept minus 0.2truept

\vfill
\mbox{Feb 2000}
\end{titlepage}

\baselineskip=21truept plus 0.2truept minus 0.2truept
\pagestyle{plain}
\pagenumbering{arabic}

\makebox[\textwidth][c]{\bf I. Introduction }

  Discovery of the Higgs boson will be one of the most momentous
 event in the history of physics. It will validate one of the
 most important ingredient of the standard model paradigm:
 the Higgs mechanism. Higgs mechanism is now taken for granted
 in even various attempts to go beyond the standard model; however
 experimental evidence for this mechanism remains tenuous at best.
 In the experiments that are expected to be carried out in the
 next decade, it is mostly at the LHC where one expects to find
 the Higgs boson, if it exists at all. However, before enough
 data is collected at the LHC, it may be possible to detect
 the Higgs boson at the Tevatron, if the Higgs boson exists
 in a specific mass domain. This mass domain is the lower 
 intermediate mass region ($\mZ < \mH < 130-140$~GeV).
  It is therefore of utmost importance
that the various signatures of the Higgs boson should  be 
studied in as realistic a way as possible in this mass region
at the Tevatron. 
 There exists a lower bound on the mass of the Higgs boson of order
$90$ GeV from the LEP [\ref{lepbound}]. It is expected to go
up to about $100$ GeV. The lightest Higgs boson in 
a class of supersymmetric theories is also expected to have a upper 
bound of the order of $160$ GeV. All these bounds suggest that the
intermediate mass region of the Higgs Boson is a very important region.
In this Communication, we focus on the lower intermediate mass 
Higgs boson and analyze a specific signature at the Tevatron. 
The Higgs boson that 
we study is the Higgs boson of the standard model. In some of the 
extensions of the standard model, the properties of some scalar
particles are similar to the standard model Higgs boson, therefore
our results will also be applicable in that situation.

The useful signatures of a particle at a particular collider
depends upon the possible production mechanisms and the decay properties
of the particle. There are a number of important mechanisms for the 
intermediate mass Higgs boson production at a hadron collider. However 
two of the mechanisms,
$\proda$ and $\prodb$, which are expected to be quite useful at
the LHC [\ref{lhc1},\ref{lhc2},\ref{tata}] are not that useful 
at the Tevatron. This is because
the cross-sections for the Higgs boson production through these
mechanisms at the Tevatron are quite small.
For example, for the Higgs boson of $100$ GeV mass, the cross-sections
at the Tevatron ($\sqrt{s} = 2$ TeV) are  about $0.54$ pb and $1.9$ x $10^{-3}$
 pb respectively for the above production mechanisms. Therefore, with
 the expected luminosity, rare decay 
 channels of the Higgs boson for $\proda$ and
 any decay channel for $\prodb$ will give rise to very few if any
 events at the Tevatron.
 In this Communication, we specifically consider a third production
 mechanism where a Higgs boson is produced in association with
 a W-boson:

\begin{equation}
  \label{eq:1}
 \signal \, .  
\end{equation}

In the lower intermediate mass region, the Higgs boson decays 
predominantly into two bottom quarks while the W-boson decays 
into two leptons, $l^\pm \nu_l$, or two quarks. 
Here `$l$' can be an electron, a muon, or a tau-lepton.
The useful signatures th[\ref{ae},\ref{smw},\ref{abc},\ref{ak}] 
through this mechanism require that
the W-boson decays leptonically.
The signature \signature due to this mechanism has been studied 
in detail using parton level Monte-Carlos as well as event
generators at the LHC. One event generators based
study suggests that a modified version of this signature could
be useful. We shall study this same signature at the Tevatron.
This signature requires the identification of the bottom jet.
Using silicon-vertex detectors, it is possible to identify
a bottom jet with an  efficiency factor of about
$40-50\%$ [\ref{kul}]. However, one also has to take into account
the possibility  of a jet due to a parton other
than the bottom quark mimicking a bottom jet. This would mean 
that even processes without a bottom jet can be a background
to the signature in which we are interested.
Therefore a study should include all such backgrounds.

  The parton level 
Monte-Carlo analyses primarily use tree level
cross-sections and distributions for the signal and the various
backgrounds. The effects of radiative corrections, hadrionization
and a lot of other details of an event are not simulated. 
Such studies are often quite useful as a first step, but in
the end to find the feasibility of a signature, a more realistic
study is needed.
 In this Communication,
we have carried out a study after including a few background
processes in the event generator PYTHIA [\ref{pythia}] to take into account
major effects of the next-to-leading-order (NLO) corrections.
It turns out that one should look at a modified version of the 
signature that has been considered until now.
 
    The unmodified version of the exclusive signature that we 
  study is \signature. Here by
  exclusive, one means that we veto any event that has extra 
  hard particles
  other than an isolated charged lepton and two bottom jets in
  a particular kinematic domain.  Therefore
  any process that can give rise to a W-boson and two bottom jets
  with the W-boson decaying leptonically is a background.
  As discussed above, another source of background is due to
  flavor misidentification. A gluon or light quark-initiated jet can
  mimic a bottom jet with a small probability; we therefore also
  consider processes that give rise to
 potential backgrounds due to this misidentification. The modification
 of this signature that we introduce is discussed below.

   The background processes are broadly speaking are of two types
 [\ref{abc}]:
 the $W$-boson-associated backgrounds and the top-quark-associated
 backgrounds. 
  The $W$-boson-associated backgrounds are:
       \begin{eqnarray}
         \label{eq:2}
         p \bar{p} \to W b \bar{b} X, \; \; W Z X, \; \; W j j X
       \end{eqnarray}

    Here the $\bgB$ is a background when the Z-boson decays
 into a pair of bottom quarks. The process $\bgC$ is background
 when a jet mimics a bottom jet. The signal process $\signal$
 is already a part of the PYTHIA package; so is the process
 $\bgB$. We have included the processes $\bgA$ and $\bgC$ in PYTHIA. 
 The inclusion of all the above processes in PYTHIA allows
 us to study the effects of part of the NLO
 corrections to the leading-order (LO) results.

       The second class of backgrounds, the top-quark-associated 
 backgrounds are:

       \begin{eqnarray}
         \label{eq:3}
          p \bar{p}  \to t \bar{t} X, \; \; t b X, \; \; t q X, \; \; t q b X.
       \end{eqnarray}

  These processes are backgrounds when the top quark decays
 into a W-boson and a bottom quark. The primary motivation
 to look for an exclusive rather than an inclusive signature
 is due to the $ p p \to t \bar{t} X $ background. This process
 always have particles other than a $e/\mu$ and bottom quarks
 in the final state. Therefore by requiring that there be
 only a few jets in the final state, we can reject large
 fraction of the top quark background. Here the process
 $\bgD$ is already in PYTHIA. The last three processes are not in 
 PYTHIA. These processes are not most important sources of the
 backgrounds at the Tevatron but are not insignificant.
 We are including these processes also in PYTHIA and shall present the 
 results in a more complete study elsewhere. More details about all the
 above backgrounds can be found in Ref [\ref{abc}].

Although we are using an event generator which provides a means
to include some of the NLO effects, but the cross-section of
a process is still normalized to the leading order result.
This type of Monte-Carlo simulation usually includes only
bremsstrahlung-type corrections and the effects of hadronization; 
inclusion of loop-corrections is not possible in most event generators
as is the case with PYTHIA. However, our interest
is in that part of NLO corrections that gives rise to extra soft jets,
for which purpose PYTHIA can serve fairly well. One can multiply
by an overall K-factor to get proper normalization 
of the cross-section. The
QCD corrections to the signal production have been
found to be about 10--15$\%$ [\ref{willenHan}, \ref{resum}].
While for the
backgrounds, the corrections will depending on the process
 These corrections could vary in the range of
10--30$\%$ [\ref{nason}]. Therefore the estimates of our
signal and background events should be increased by about
10--15$\%$  on average. Since at the Tevatron the number 
of signal events is at premium, therefore a proper appraisal
of a signature should include this factor.


The paper is organized as follows: In the next section, we describe
how the parton level results change due to the inclusion
the NLO effects which are normally included in an event generator. In
Sec. III, we discuss the numerical results for the
various processes. In this way, we assess the
usefulness of the modified exclusive \signature signature. 
 In Sec.IV, we present our conclusions.

\vspace{0.4cm}

\makebox[\textwidth][c]{\bf II. Effects in Event Generators}

\vspace{0.4cm}

    In our analysis, we have used the event generator PYTHIA
  to take into account some of the NLO corrections.
  Such event generators have three new ingredients:1) the initial-state
  radiation (ISR), 2) the final-state radiation (FSR), and 
  3) hadronization and decays (H \& D). As is obvious from
  the terminology, the initial-state radiation corresponds
  to the emission of the partons or leptons from the initial-state
  particles; the final-state radiation is similar emission
  from the final-state particles; hadronization and decays
  corresponds to the hadronization of the partons and subsequent
  decays of the hadrons.

   We can illustrate the effects of the new ingredients
  on the parton level results by considering the signal. 
  In PYTHIA, we can turn-off all these effects and
  turn-on one new effect at a time to assess its impact on the
  leading order results.
  We are starting with the process $\signal$. At the leading order,
  at the parton level,
  we have an electron and two bottom quarks in the final state (apart
  from the partons that do not take part in the hard scattering). 
  When we turn-on the
  initial-state radiation, it changes the structure of events.
  A large fraction of events will have more than two hard partons 
  (bottom quarks)
  in the final state. Therefore, even with a minimum $p_T$ cut
  on the partons, we shall have a significant numbers of events
  with more than two jets in the final state. At the Tevatron,
  \eg, about $40-50\%$ of events will have one or more extra jets
  (other than two bottom jets)  with a $p_T$ cut of about $15$
  GeV. Therefore, there will be a reduction in the number of signal
  events, as compared to the leading order, if we consider the
  exclusive signature \signature. One obvious way to get around this 
  problem is to broaden the signature to include even events
  with one or more extra jets. The negative aspect of this new 
  signature is
  that there will be a larger background from processes that
  have more than two jets without any initial state radiation,
  \eg,  $\bgD$. Luckily at the Tevatron the top-quark-associated
  backgrounds are not as dominant as the W-boson-associated backgrounds.
  The impact of the final-state radiation is similar to that
  of initial-state radiation with a few differences. Now
  suppose there is only final-state radiation. There will
  be fewer events with extra hard jets as compared to the initial-state
  radiation; this is because of the smaller scale of radiation.
  However now the mass of two
  bottom-jet system will be broader and the peak of
  the distribution will be shifted to a lower value
  than without the final-state radiation. This is because 
  the final-state radiation would come from bottom quarks 
  (which in turn are there because of the decay of the Higgs boson) 
  and therefore it will drive the $M(bb)$ peak away from the $m_H$
  mass. Since W-boson-associated background will increase due to this, 
  the final-state radiation leads to increase in the overall background.
  One can attempt to improve upon the situation by including
  extra jets in the mass distribution; so one may wish
  to make a cut on $M(bbj)$ or $M(bbjj)$ distributions also.
  The effect of the
  hadronization and decays is also to dilute the parton-level results.
  This also shows up as pulling the $M(bb)$ distribution peak
  below the Higgs boson mass and broadening of the distribution.
  Lowering of the distribution peak enhances the W-boson-associated
  background; while a broader distribution increases all backgrounds
  if we wish to maintain a large number of signal events.

   Because of above changes the {\it modified} signature that we 
  shall consider will be to also include events that have
  one extra jet (other than two bottom jets). This
  will increase the number of signal events, but also the
  total background events. But since it is important to enhance
  the number of signal events, we use this modified signature. 
  One can consider more than one extra jet also, but it will lead
  to significant increase in the top-quark associated backgrounds.

\vspace{0.4cm}

\makebox[\textwidth][c]{\bf III. Numerical Results}

\vspace{0.4cm}

         In this Section, we present an estimate of the
signal and background events rates at the Tevatron. We 
take the center of mass energy to be $2$ TeV. 
If we apply the following typical acceptance cuts,

\begin{eqnarray}
  \label{eq:4}
& &  p_{T}^{\ell} > 10 \;\; {\rm GeV}; \;\;\;\;\;\; 
  p_{T}^{b,j} > 15 \;\; {\rm GeV}; \;\;\;\;\;\; 
  |\eta|^{\ell, b, j} < 2.5; \nonumber \\
 & & \;\;\;\;\;\;\;\;\;\;\;  \Delta R(\ell, b) > 0.7;
   \;\;\;\;\;\;  \Delta R(b, b) > 0.6;
\end{eqnarray}

   where the index $\ell$ stands for either electron or muon; 
 $\eta$ is pseudo-rapidity and
 $\Delta R = \sqrt{(\Delta \eta)^2 + (\Delta \phi)^2 }$, we find that
the approximate cross-sections are: $\sigma (\signal) = 0.2$
pb ($m_H = 100$ GeV); $\sigma (\bgA) = 5.2$ pb; $\sigma (\bgB) = 2.6$ 
pb; $\sigma (\bgC) = 660.1$ pb; $\sigma (\bgD) = 6.4$ pb.
As we clearly see that combined cross-section for the backgrounds
is three-four orders of magnitude higher than the signal.
However, the $\bgC$ cross-section, when multiplied by the mimic
probability and other cross-sections when multiplied by
appropriate branching ratios, we find that the combined
background is about two orders of magnitude higher than
the signal.To reduce that background further, the useful
physical quantities are: $p_T^{b,j}$, $\Delta R(b,b)$, 
$M(bb)$, cos($\Delta \phi$),
cos$^{REC}_{CM}(H)$. Here $M(bb)$ is the mass of two-bottom-jets
system; cos($\Delta \phi$) is cosine of the difference of the
azimuthal angle of the two bottom quarks; cos$^{REC}_{CM}(H)$ is
the angle of the two bottom quark system with the z-axis in 
the reconstructed
center-of-mass frame. Because of the existence of a neutrino and
soft particles in the final state, we cannot go the center-of-mass 
frame.
However, by deducing the unknown longitudinal momentum of the
unobserved particles, as discussed in Ref [\ref{abc}], it is possible to
reconstruct approximate center-of-mass system.

    The observable $M(bb)$ is most effective. Therefore we shall first
 consider only acceptance cuts and a cut on the $M(bb)$ observable. We
 shall briefly discuss the usefulness of other observables towards
 the end of this section. As we discussed in the last section,
 the final-state radiation and hadronization $\&$ decays shift 
 the $M(bb)$ peak below the $m_H$ mass
 for the $\signal$ process. Therefore to include most of the signal,
 we look for events around the shifted peak that
 is approximately around $m_{peak} = m_H - 20$ GeV. We are displaying
 the results for $|M(bb) - m_{peak}| < 10$ and $15$ GeV in Tables
 $1$ and $2$ respectively. The smaller
 value of 10 GeV will reduce the background, but it will also reduce
 the signal. Since at the Tevatron, the number of signal events is
 quite small, therefore it may be necessary to choose a value of
 15 GeV for this cut. To arrive at the numbers in the tables
 we have taken the mimic probability to be one percent and
 the bottom-jet identification efficiency to be about $45\%$.
 The extra-jet can have $p_T > 8$ GeV and $\Delta R(b,b) > 1.2$; rest 
 of the cuts are as in equation 4.
 Here we have taken the integrated
 luminosity to be 10 fb$^{-1}$, which may take several years for
 the Tevatron to accumulate, until the Tevatron start operating
 in high luminosity mode. However,
 we see that one may need about 20 fb$^{-1}$ before one may be able
 to detect the Higgs boson in the lower-intermediate-mass region.
 Only with such integrated luminosity, one may be able to achieve
 a significance $(S/ \sqrt{B})$ value of about 3.

  Since one of the problem at the Tevatron is to have enough signal
 events, we would like to comment briefly at the maximum possible
 number of events for the signature under consideration. We would 
 like to make several points: a) When the W-boson decays into a 
 $\tau$-lepton and it subsequently decays about $35\%$ of the time
 into a $e/\mu$, this decay-chain will contribute to the signal
 events (and also similarly to the background); since the
 electron/muon from a decay chain will be somewhat softer,
 conservatively, we may expect that W-boson decaying into a tau-lepton
 will contribute about $25\%$ of the events as compared to when 
 the W-boson decays directly into an electron or a muon; b) The
 cross-sections that we have used are leading-order cross-sections;
 It is known that the NLO corrections will enhance the signal at
 most by about $15\%$ (the backgrounds will also be enhanced by about
 the same order); c) We have not tried to find the optimum set of
 cuts; if one tries and uses neural-net or decision-tree like 
 techniques, the signal may still be further enhanced by $15-20\%$;
 d) the bottom-jet identification efficiency could be as
 high as about $50\%$ and one may be able to reconstruct the jet
 better; this may increase the event rate by $10-15\%$; e)
 one may include events with two or more extra jets; but this
 will increase the top-quark-associated backgrounds a lot.
 In Table 3 we display the estimate of the signal and the backgrounds
 including the a) and b) enhancements (potential enhancement c) and
 d) and e) are not included).

   From these tables we notice that the most troubling background
 is $\bgA$. This is so large in part because of the peak-shift and
 broadening of the $M(bb)$ distributions for the signal. The 
observables cos($\Delta \phi$) and cos$^{REC}_{CM}(H)$ can help 
us to reduce this background.($\Delta R(b,b)$ and $p_T^{b,j}$
are also useful. But their effects can be taken care by cos($\Delta
\phi$) and $M(bb)$ distributions respectively.)
 In Ref [\ref{abc}] these distributions were
discussed. We find that it is possible to reduce the $\bgA$ background
by $50-70\%$ with a reduction in the signal events of about $25-30\%$.
However, it does not enhance the significance by a large factor
because of small number of signal events to begin with. A more
complete study is required to put these and other such observables
to optimum use.

\vspace{0.4cm}

\makebox[\textwidth][c]{\bf IV. Discussion and Conclusions}

\vspace{0.4cm}

    In this Communication, we have studied the {\it modified} 
 exclusive signature
 \signature for the associated production of the lower-intermediate-mass
  Higgs boson. This modified signature includes an extra jet.
Such a signature results
 through the process $pp\to WHX \to \ell\nu_{\ell} b\bar bX$ when
 we include the NLO corrections. Our
 study is based on the PYTHIA event generator. We have
 considered the principal backgrounds and found that with enough
 luminosity one should be able to detect the Higgs boson if it
 exists in the studied mass region. At the Tevatron, the dominant
 background is due to $\bgA$. However it can be controlled
 with judicious choice of cuts. We have not attempted to find 
 optimum set of cuts; however one can use the observables discussed
 here and in the Ref [\ref{abc}] to reduce the background further. As we see
 that there is a need to enhance the number of signal events. 
 By modifying the cuts on the observables, one may be able to enhance the signal
 by at most extra $20-30\%$ as compared to the numbers given in the
 Table 3. However the background will also 
 increase and one will have to strike a balance to enhance the
 significance of the signature.
 The accumulated integrated luminosity will be critical
 for the detection of the Higgs boson. It would appear that one
 will need about 20 fb$^{-1}$ of accumulated luminosity before something
 definite could be said about the existence of the Higgs boson
 in the lower intermediate mass region. Such a luminosity may be
 reached if the Tevatron runs in high luminosity mode for several years.
 A more complete study is underway.

\bigskip
\bigskip
\bigskip

\noindent{\large ACKNOWLEDGEMENTS}

\medskip

I would like to thank Gobinda Kar for his participation in
the earlier stages of Higgs-boson related projects.

\vskip .5in

\relax
\def\pl#1#2#3{
     {\it Phys.~Lett.~}{\bf B#1} (#3) #2}
\def\mpla#1#2#3{
     {\it Mod.~Phys.~Lett.~}{\bf B#1} (#3) #2}

\def\zp#1#2#3{
     {\it Zeit.~Phys.~}{\bf C#1} (#3) #2}
\def\prl#1#2#3{
     {\it Phys.~Rev.~Lett.~}{\bf #1} (#3) #2}
\def\rmp#1#2#3{
     {\it Rev.~Mod.~Phys.~}{\bf #1} (#3) #2}
\def\prep#1#2#3{
     {\it Phys.~Rep.~}{\bf #1} (#3) #2}
\def\pr#1#2#3{
     {\it Phys.~Rev.~ }{\bf D#1} (#3) #2}
\def\np#1#2#3{
     {\it Nucl.~Phys.~}{\bf B#1} (#3) #2}
\def\ib#1#2#3{
     {\it ibid.~}{\bf #1} (#3) #2}
\def\nat#1#2#3{
     {\it Nature (London) }{\bf #1} (#3) #2}
\def\ap#1#2#3{
     {\it Ann.~Phys.~(NY) }{\bf #1} (#3) #2}
\def\sj#1#2#3{
     {\it Sov.~J.~Nucl.~Phys.~}{\bf #1} (#3) #2}
\def\ar#1#2#3{
     {\it Ann.~Rev.~Nucl.~Part.~Sci.~}{\bf #1} (#3) #2}
\def\ijmp#1#2#3{
     {\it Int.~J.~Mod.~Phys.~}{\bf #1} (#3) #2}
\def\cpc#1#2#3{
     {\it Computer Physics Commun. }{\bf #1} (#3) #2}

\begin{reflist}
\item \label{lepbound} S. Banerjee, Talk presented at WHEPP5, Pune, India (Janaury 199
8).

\item \label{lhc1} ATLAS Technical Proposal, CERN/LHCC 94-43 (December 1994).

\item \label{lhc2} CMS Technical Proposal, CERN/LHCC 94-38 (December 1994).

\item \label{tata} P. Agrawal, \pl{229}{145}{89}.

\item \label{ae} P. Agrawal and S. Ellis, \pl{229}{145}{1989}.

\item \label{smw} A. Stange, W. Marciano, and S. Willenbrock,
\pr{49}{1354 }{1994}; \pr{50}{4491}{1994}.

\item \label{abc} P. Agrawal, D. Bowser-Chao and  K. Cheung, \pr{51}{6114}{1995}.

\item \label{ak} P. Agrawal and M. Kar, IOP preprint IP/BBSR/97-31.

\item \label{kul} Report of the tev\_2000 Study Group, 
                            Fermilab-Pub-96/082 (April 1996).

\item \label{pythia}  T. Sjostrand, {\it Computer Physics Commun.} 82 (1994) 74.

\item \label{willenHan} T. Han and S. Willenbrock, \pl{273}{167}{91};
J. Ohnemus and W.J. Stirling, \pr{47}{2722}{93};
 H. Baer, B. Bailey, and J.F. Owens,\pr{47}{2730}{93}.

\item \label{resum} P. Agrawal, J. Qiu, and C.  P. Yuan, MSU-HEP-93-8;
P. Agrawal, J. Qiu, and C.  P. Yuan, Proceedings of the Workshop on
{\em Physics at Current Accelerators and Supercolliders}, Argonne
National Laboratory (June 2--5, 1993).

\item \label{nason}P. Nason, S. Dawson, and R.K. Ellis,  \np{303}{607}{88};
W. Beenakker, H. Kuijf, W. L. Neerven and J. Smith, \pr{40}{54}{89}.

\end{reflist}

\newpage



\begin{tabular}{||c|c|c|c|c|c||} \hline
            &  $\mH $ &  $\mH$  &  $\mH$   &  $\mH $ & $\mH$ \\
  Processes &  90 GeV & 100 GeV &  110 GeV & 120 GeV & 130 GeV \\ \hline
$WH$   &  32 & 23  & 16 & 11 & 6  \\
$Wjj$  &  12 & 10  &8  & 7  & 6 \\
$Wbb$  & 160 & 130 & 99  & 75 & 54 \\
$WZ$   &  41 & 32 & 15 & 6 & 3 \\ 
$t \bar{t}$ &17  &  15   & 15 & 15  & 15  \\ \hline
\end{tabular}
\vskip .1in
{\small
Table 1. Event rates at the Tevatron with 10 fb$^{-1}$ of 
accumulated integrated luminosity with the acceptance cuts
given in the text and $|M(bb) - m_{peak}| < 10$ GeV.
}
\vskip .3in
\hskip -.5in

\begin{tabular}{||c|c|c|c|c|c||} \hline
            &  $\mH $ &  $\mH$  &  $\mH$   &  $\mH $ & $\mH$ \\
  Processes &  90 GeV & 100 GeV &  110 GeV & 120 GeV & 130 GeV \\ \hline
$WH$   &  39 & 29 & 21 & 14 & 8  \\
$Wjj$  &  16 & 13 & 12 & 10 & 8 \\
$Wbb$  &  211  & 170  & 140 & 112 & 91 \\
$WZ$   &  52    & 42 & 25 & 10 & 4 \\ 
$t \bar{t}$ & 23  & 25 & 23 &  23&  23  \\ \hline
\end{tabular}
\vskip .1in
{\small
Table 2. Event rates at the Tevatron with 10 fb$^{-1}$ of 
accumulated integrated luminosity with the acceptance cuts
given in the text and $|M(bb) - m_{peak}| < 15$ GeV.
}

\vskip .3in
\hskip -.5in

\begin{tabular}{||c|c|c|c|c|c||} \hline
            &  $\mH $ &  $\mH$  &  $\mH$   &  $\mH $ & $\mH$ \\
  Processes &  90 GeV & 100 GeV &  110 GeV & 120 GeV & 130 GeV \\ \hline
$WH$   &  51 & 38 & 27 & 18 & 11 \\
$Wjj$  &  21 & 17 & 14 & 13 & 11 \\
$Wbb$  &  271     & 221 & 182 & 146 & 117 \\
$WZ$   &  67    & 55 &  32 & 13 & 4 \\ 
$t \bar{t}$ & 30  & 32    & 30 & 30 & 30   \\ \hline
\end{tabular}
\vskip .1in
{\small
Table 3. Event rates at the Tevatron with 10 fb$^{-1}$ of 
accumulated integrated luminosity with the acceptance cuts
given in the text and $|M(bb) - m_{peak}| < 15$ GeV. The number
also include the enhancement due to tau-lepton decays and
the NLO K-factor.
}

\end{document}